\documentclass[12pt,preprint]{aastex}

\slugcomment{Preprint DAMTP-2001-55; July 12th, 2001}
\shorttitle{A Supernova Brane Scan}
\shortauthors{P.P. Avelino and C.J.A.P. Martins}

\begin{document}

\title{A Supernova Brane Scan}

\author{P.P. Avelino\altaffilmark{1} and C.J.A.P. Martins\altaffilmark{2,3}}
\affil{Centro de Astrof\'{\i}sica da Universidade do Porto,\\
Rua das Estrelas s/n, 4150-762 Porto, Portugal}

\altaffiltext{1}{Dep. de F{\' \i}sica da Faculdade de Ci\^encias da
Univ. do Porto, Rua do Campo Alegre 687, 4169-007 Porto, Portugal.
Email: pedro\,@\,astro.up.pt}
\altaffiltext{2}{Department of Applied Mathematics and Theoretical Physics,
Centre for Mathematical Sciences, University of Cambridge,
Wilberforce Road, Cambridge CB3 0WA, United Kingdom.
Email: C.J.A.P.Martins\,@\,damtp.cam.ac.uk}
\altaffiltext{3}{Institut d'Astrophysique de Paris,
98 bis Boulevard Arago, 75014 Paris, France}

\begin{abstract}
We consider a `brane-world scenario' recently introduced by Dvali,
Gabadadze and Porrati, and subsequently proposed as an alternative
to a cosmological constant in explaining the current acceleration of
the universe. We show that, contrary to these claims, this particular
proposal is already strongly disfavoured by the available Type Ia
Supernovae, Cosmic Microwave Background and cluster data.
\end{abstract}

\keywords{Cosmological parameters---cosmology: theory---methods: data
analysis---supernovae: general}

\section{Introduction}

At a time when observational cosmologists are finally pinning down some
crucial cosmological parameters
\citep{perl1,perl2,riess,boom1,boom2,boom3,maxi1,maxi2,maxi3,dasi1,dasi2,tdf},
theoretical cosmologists have increased and diversified their efforts
to try to provide some more solid connections between particle
physics and cosmology.

In this context, a topic of much recent interest has been that of the
so-called `brane-world scenarios' \citep{orig1,orig2,orig3,review1,review2}.
The general principle behind such models is that
the ordinary particles live on a three-dimensional surface (commonly called
a 3-brane, or simply `the brane'), which is embedded in a larger space
(`the bulk', which may or may not be compact and might even have an
infinite volume) on which gravity can propagate. An observer on the brane
will measure four-dimensional gravity up to some corrections which,
given the weakness of gravity, can in general be made small enough not to
conflict with observations without tweaking with model parameters too much.

At present the topic is young enough that the main drive is still to
try to explore all remotely viable model-building possibilities without
worrying too much about the consequences. However, some of the proposed
models are already developed enough that they can start to be put
through the sieve of specific cosmological observations. In this paper
we will provide what we believe to be the first detailed analysis of
this kind for a brane world scenario.

We will consider a particular solution of a brane world scenario
originally introduced by \citet{dgp},
and further studied in \citep{def,ddg}---we shall refer to it as the
DGP model for simplicity. It's a five-dimensional brane-world model
with a non-compact, infinite-volume extra dimension. The usual
four-dimensional gravity is recovered on the brane for scales
below a `characteristic radius' $r_c$, due to a four-dimensional Ricci
scalar being induced on the brane. However, at larger scales this
becomes sub-dominant, and one will effectively see five-dimensional
Einstein-Hilbert gravity.

The above effect can obviously have dramatic cosmological implications.
A particularly interesting solution was first found
by \citet{def}---and then generalized by \citet{rdick}---and
then further studied in \citep{ddg}. It describes a universe which
at late times is accelerated on scales larger than $r_c$. This is an effect
of the bulk gravity, in the sense that
observers on the brane will see no cosmological
constant. Hence this is another interesting alternative way to explain the
current acceleration of the universe, which is strongly indicated by
Type Ia supernova observations \citep{perl1,perl2,riess}, without
resorting to a cosmological constant---for earlier alternative explanations,
see \citep{dev,behnke,mann}. Note that, unlike most other known
brane worlds scenarios, here the early evolution of the universe is 
the standard one while the late evolution is different. Also, unlike other
alternative theories of gravity (introduced in other contexts) here
gravity will become {\em weaker} on large enough scales. These two
points will be important in what follows.

In \citep{ddg} the authors argue that this is an intrinsically
higher-dimensional effect, at least in the sense that one can not mimic
it with arbitrary high-derivative terms in ordinary four-dimensional
gravity. This turns out to be both a blessing and a
curse, for on one hand it means that one can extract quite distinctive
observational predictions, but on the other hand it
also implies that it's
quite easy to rule it out. In \citep{ddg} the authors claim that the model's
alternative explanation for the current acceleration of the universe
agrees with all existing cosmological observations (or, more accurately, that
it is currently indistinguishable from the standard scenario). In what
follows we shall show that this is not the case. Indeed, currently
existing data is already sufficient to make this alternative
explanation for acceleration strongly
disfavoured when compared to the standard one.

The plan of the paper is as follows. In Sect. 2 we provide
a brief summary of the features of the DGP model which are relevant for
our discussion. We then proceed to analyse the accelerating
solution in the light of
the Type Ia supernovae data in Sect. 3. In Sect. 4
we cross-check the results of this analysis with other cosmological
datasets, and finally we present our conclusions in Sect. 5.

\section{The Model}

Here we briefly describe the brane-world scenario 
introduced by \citet{dgp},
and further studied in \citep{def,ddg}---we shall henceforth
refer to it as the DGP model. Our discussion will be somewhat
simplified, as we will only focus on the features that are relevant for
our subsequent analysis---the reader is encouraged to consult the
original references for a more detailed discussion.

Our three-brane is embedded in a five-dimensional spacetime
with a non-compact, infinite-volume extra dimension. Particles in
the standard model are confined to the brane, and brane fluctuations are
neglected. There is essentially one free parameter in the model,
which is the `five-dimensional Planck mass', denoted $M_5$. Note
that one {\em must} assume that the standard model cut-off doesn't
coincide with $M_5$---in fact, it must be much larger, so that the
physical interpretation of $M_5$ is not quite trivial.

The four-dimensional Planck mass will be denoted $M_4$, and is related
to the usual gravitational constant through $8\pi G_4=M_4^{-2}$. Unlike
in other brane world scenarios, here the two masses $M_4$ and $M_5$
need not be related. We note that there is a somewhat technical problem
with the model \citep{def} which implies that if one defines Newton's
constant via a standard Cavendish-like experiment, then the so-defined
$G_N$ doesn't necessarily coincide with $G_4$. This would obviously
contradict standard tests of General Relativity, as was already
pointed out in \citep{def}. Possible ways to circumvent this problem have
been claimed \citep{dgp,ddg}. In any case,
we shall neglect this aspect in what follows, since our present purpose
is to discuss `cosmological' (as opposed to `local') tests of the model.

The usual four-dimensional gravity is recovered on the brane for scales
below a `characteristic radius' $r_c$, which is given by
\begin{equation}
\label{defrc}
r_c\equiv\frac{M_4^2}{2M_5^3}\,.
\end{equation}
This is due to a four-dimensional Ricci
scalar being induced on the brane. However, at larger scales this
term becomes sub-dominant, and one will effectively see five-dimensional
Einstein-Hilbert gravity. Therefore gravity becomes {\em weaker} on large
enough scales. This is to be contrasted with models where one modifies gravity
on large scales in order to solve, for example, the dark matter problem: in
that context, one requires stronger gravity on large scales.
From this it immediately follows that one can impose
a simple constraint on $M_5$, since the characteristic radius must at
least be as large as the present Hubble radius.

Cosmological solutions in this model were first studied
in \citep{def}---but see also \citep{rdick}.
One finds that the Friedmann equation on the brane has the following form
\begin{equation}
\label{friedmann}
H^2+\frac{k}{a^2}=\left(\left[\frac{8\pi}{3}G_4\rho+
\frac{1}{4r_c^2}\right]^{1/2}+\frac{\epsilon}{2r_c}\right)^2\,,
\end{equation}
where $k=0,\pm1$ is the spatial curvature and $\epsilon=\pm1$ corresponds
to two different brane embeddings in the bulk spacetime.
On the other hand, the energy conservation equation has the standard form,
\begin{equation}
\label{energycons}
\frac{d\rho}{dt}+3H(\rho+p)=0\,.
\end{equation}

At early enough times the density term dominates the Friedmann equation,
and hence one obtains (at least to first order) the standard cosmological
evolution, namely $3H^2=8\pi G_4\rho$.
The additional bulk-induced term will become important when
$H^{-1}\sim r_c$. Then the subsequent evolution depends on the sign of
the parameter $\epsilon$. In the $\epsilon=-1$ branch the universe
switches into a full five-dimensional gravity regime, where the
Friedmann equation looks like $H\propto\rho$---something that is typical
of many brane world scenarios. On the other hand, in the
$\epsilon=+1$ branch something rather more interesting happens. There is
a `self-inflationary' solution with $H\sim r_c^{-1}$. What happens is
that the additional curvature term on the brane appears as a source for bulk
gravity, and thus can cause acceleration on the brane. In other words,
an observer on the brane will see the universe being 
accelerated on scales larger than $r_c$.

Note, however, that this solution does not require any other energy source
on the brane---so in this sense this is indeed a higher-dimensional effect.
In particular, no cosmological constant is needed on the brane, so this is
an interesting alternative way to explain the
current acceleration of the universe, which is strongly indicated by
Type Ia supernova observations \citep{perl1,perl2,riess}, without
resorting to a cosmological constant. A simple, `back-of-the-envelope'
constraint comes from the fact that we want the universe to be at this 
crossover stage at about the present epoch if the alternative proposal
for the acceleration of the universe is to be viable,
hence $H_0^{-1}\sim r_c$. This then naively
implies that the five-dimensional Planck mass should be of the order of
\begin{equation}
\label{gconstaccel}
M_5\sim10-100MeV\,.
\end{equation}

We finally note that high-energy processes place almost no constraints on
this mass scale $M_5$, basically because up to about the present epoch the
universe evolves as normal. Indeed the only constraint comes from the
measurement of the Newtonian force, which only implies the very
mild
\begin{equation}
\label{gconst}
M_5>10^{-3}eV\,.
\end{equation}
As we shall see in the following section, much more stringent constraints
can be derived using cosmological observations, if one assumes that the
accelerating solution is valid. We will start by an analysis
of the Type Ia supernovae data, which is described below. We will
then contrast the results of this analysis with other cosmological
constraints.

\section{Supernovae Data Analysis}

We begin by evaluating the luminosity distance as a function of 
the cosmological parameters for our model. The Friedmann equation 
(\ref{friedmann}) can be rewritten as a function of 
the red-shift $1+z \equiv a_0/a$ to give: 
\begin{equation}
\label{omegas}
\frac{H^2(z)}{H_0^2}=\Omega_k(1+z)^2+\left(\sqrt{\Omega_{rc}}+
\sqrt{\Omega_{rc}+\sum_\alpha\Omega_\alpha(1+z)^{3(1+w_\alpha)}}\right)^2\,,
\end{equation}
where 
\begin{equation}
\label{omegacurv}
\Omega_k=\frac{-k}{H_0^2a_0^2}, \qquad \qquad \Omega_{rc}=
\frac{1}{4r_c^2H_0^2}, \qquad \qquad \Omega_\alpha=
\frac{8\pi G_4\rho_{\alpha0}} {3H_0^2a_0^{3(1+w_\alpha)}},\,
\end{equation}
respectively represent the fractional contribution of curvature,
the bulk-induced term and the other components in the 
Friedmann equation. In equation (\ref{omegas}) the sum is over 
all the components of the cosmic fluid with an equation of state 
$p_\alpha = w_\alpha \rho_\alpha$. From 
now on we will consider only one component of non-relativistic particles
together with the bulk-induced term, 
in which case equation (\ref{omegas}) becomes
\begin{equation}
\label{omegas1}
\frac{H^2(z)}{H_0^2}=\Omega_k(1+z)^2+\left(\sqrt{\Omega_{rc}}+
\sqrt{\Omega_{rc}+\Omega_m (1+z)^3}\right)^2\,;
\end{equation}
in particular, at the present day one must have
\begin{equation}
\label{normaliztoday}
\Omega_k+\left(\sqrt{\Omega_{rc}}+\sqrt{\Omega_{rc}+\Omega_m}\right)^2=1\,.
\end{equation}
For a flat universe $\Omega_k=0$ and so in this case the two
cosmological parameters are related by
(\ref{normaliztoday}) 
\begin{equation}
\label{flatcase}
\Omega_{rc}=\left(\frac{1-\Omega_m}{2}\right)^2\,.
\end{equation}

It is also possible to show that closed universes with
\begin{equation}
\label{bouncing}
|\Omega_k|^{3/2} > 8 \Omega_m \Omega_{rc}^{1/2}\,
\end{equation}
do not have a big bang. These universes avoid the big bang singularity
by bouncing in the past. We shall disregard such universes in most of
what follows.
Another useful benchmark is the redshift at which the universe
switches from deceleration to acceleration, or in other words the
redshift for which the deceleration parameter vanishes.
For a flat universe $\Omega_k=0$ it's easy to show that the following
exact result holds
\begin{equation}
\label{qvanishes}
\left(1+z\right)_{q=0}=2\left(\frac{\Omega_{rc}}{\Omega_m}\right)^{1/3}\,;
\end{equation}
note that in this (flat) case $\Omega_{rc}$ and $\Omega_m$ are not
independent parameters---they are related by (\ref{flatcase}). For comparison,
the result in the standard case, again taking a flat model ($\Omega_m
+\Omega_\Lambda=1$) is
\begin{equation}
\label{usualqvanishes}
\left(1+z\right)_{q=0}=\left(\frac{2\Omega_\Lambda}{\Omega_m}\right)^{1/3}\,;
\end{equation}
we shall return to these quantities in the following section.

It is also straightforward to show that in a Friedmann-Robertson Walker (FRW) 
universe the luminosity distance is given by:
\begin{equation}
\label{ldistance}
d_L=H_0^{-1}(1+z)|\Omega_k|^{-1/2} S_k(|\Omega_k|^{1/2} d_C)\,,
\end{equation}
where
\begin{equation}
\label{cdistance}
d_C=H_0\int_0^z \frac{dx}{H(x)}\,,
\end{equation}
and $S_k$ is $\sinh$ if $\Omega_k \ge 0$ and $\sin$ if $\Omega_k < 0$.
Note that the assumption of a FRW universe is the only one
needed to derive (\ref{ldistance}). Specifically, we {\em do not} need to
assume the validity of General Relativity or to specify {\it a
priori} the detailed contents of the universe. This is 
of course crucial in our case. 

We estimated the cosmological parameters using the combined data of two 
independent teams---thus making up a dataset of 92 different
supernovae---using the procedure described in \citep{wang,garn}. 
The measured distance modulus for a SN Ia is
\begin{equation}
\mu_0^{(l)}= \mu_p^{(l)}+\epsilon^{(l)}
\end{equation}
where $\mu_p^{(l)}$ is the theoretical prediction
\begin{equation}
\label{eq:mu0p}
\mu_p^{(l)}= 5\,\log\left( \frac{ d_L(z_l)}{\mbox{Mpc}} \right)+25,
\end{equation}
and $\epsilon^{(l)}$ is the uncertainty in the measurement, including
observational errors and intrinsic scatters in the SN Ia absolute
magnitudes. 

We denote the parameters to be fitted as {\bf s} and 
estimate them using a $\chi^2$ statistic, with \citep{riess}
\begin{equation}
\chi^2(\mbox{\bf s})=
\sum_l \frac{ \left[ \mu^{(l)}_p(z_l| \mbox{\bf s})-
\mu_0^{(l)} \right]^2 }{\sigma_{\mu_0,l}^2 +\sigma_{mz,l}^2}
\equiv \sum_l \frac{ \left[ \mu^{(l)}_p(z_l| \mbox{\bf s})-
\mu_0^{(l)} \right]^2 }{\sigma_l^2 },
\end{equation}
where $\sigma_{\mu_0}$ is the estimated measurement error of the distance
modulus, and $\sigma_{mz}$ is the dispersion in the distance modulus 
due to the dispersion in galaxy redshift, $\sigma_z$, due to
peculiar velocities and uncertainty in the galaxy redshift.
The probability density function (PDF) for the parameters {\bf s} is
\begin{equation}
\label{pppp}
p(\mbox{\bf s}) \propto \exp\left( - \frac{\chi^2}{2} \right)\,.
\end{equation}
The normalized PDF is obtained by dividing the above expression
by its sum over all possible values of the parameters {\bf s}.
In the particular case of our model the cosmological parameters are 
$H_0$, $\Omega_m$ and $\Omega_{rc}$. The probability distribution function 
for the parameters $\Omega_m$ and $\Omega_{rc}$ is obtained by integrating 
over all possible values of $H_0$, and the results are displayed
in Fig. \ref{fig1}.

We have also studied the improvements in the parameter estimation using 
supernovae which are expected from future studies of cosmic acceleration. 
Following \citep{weller} we assumed a future dataset similar to one 
proposed for the SNAP satellite\footnote{SNAP home page at
\url{http://snap.lbl.gov}}.
This has the magnitudes of 50, 1800, 50 
and 15 supernovae in the red-shift ranges from $z=0-0.2$, $z=0.2-1.2$, 
$z=1.2-1.4$ and $z=1.4-1.7$ respectively. The statistical error in 
magnitude is assumed to be $\sigma=0.15$ including both the estimated 
measurement error of the distance modulus and the dispersion in the 
distance modulus due to the dispersion in galaxy redshift. The supernovae 
dataset was generated assuming that we live in 
a standard FRW universe with cosmological parameters 
$\Omega_m=0.3$ and $\Omega_\Lambda=0.7$. Fig. \ref{fig2} shows 
the corresponding results.

\section{Results and Discussion}

\begin{figure}
\plotone{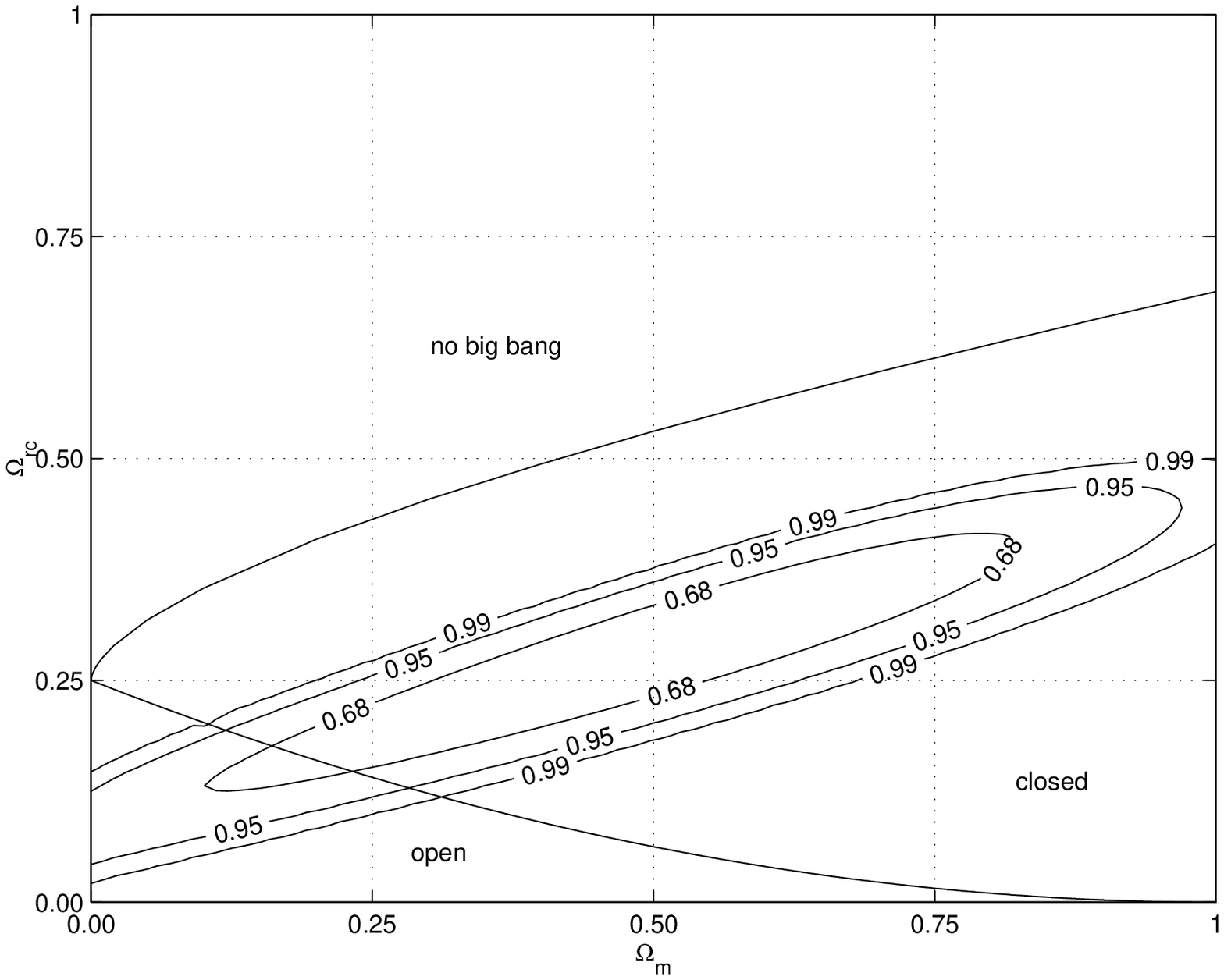}
\caption{The probability distribution function for the parameters
$\Omega_m$ and $\Omega_{rc}$ in the DGP model, for the presently
available dataset of 92 Type Ia supernovae---see the text for
detailed description of the method. The $68 \%$, $95 \%$ and $99 \%$ 
confidence contours in the $\Omega_{rc}-\Omega_m$ plane are shown, 
as well as the line separating closed universes with big bang 
from no-big-bang `bouncing' ones and the one separating closed and
open universes (ie, denoting the flat ones). \label{fig1}}
\end{figure}

\begin{figure}
\plotone{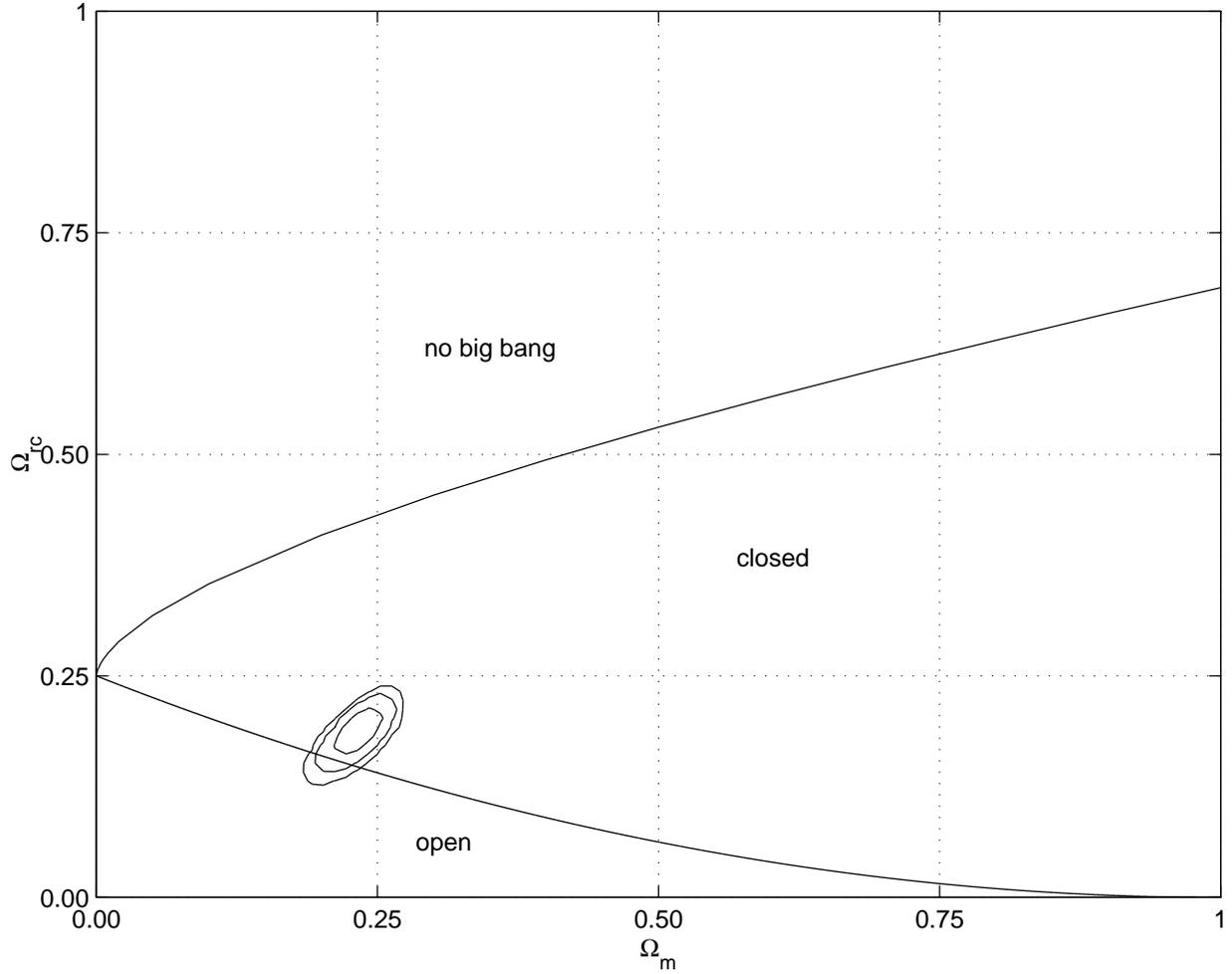}
\caption{A simulation of the analysis with a SNAP-class dataset 
generated assuming that we live in a standard FRW universe with 
cosmological parameters $\Omega_m=0.3$ and $\Omega_\Lambda=0.7$---see
the text for the other assumptions involved. Contours and boundary lines
are as in Fig. 1. \label{fig2}}
\end{figure}

A number of interesting features are apparent from Fig. \ref{fig1}.
Firstly, the likelihood analysis of the supernovae data is degenerate
in the $\Omega_{rc}-\Omega_m$ plane, approximately following a line
of the form
\begin{equation}
\label{ourdegeneracy}
\Omega_{rc}\sim \frac{2}{5}\Omega_m+\frac{1}{10}\,.
\end{equation}
This is to be compared with the standard cosmological scenario, where
the degeneracy is approximately along
\begin{equation}
\label{usualdegeneracy}
\Omega_{\Lambda}\sim \frac{4}{3}\Omega_m+\frac{1}{3}\,.
\end{equation}
Hence one can say that for any given value of $\Omega_m$,
the value of $\Omega_{rc}$ which provides the best-fit to the supernova
data is always lower than the corresponding value of $\Omega_{\Lambda}$. 
Reversing the argument, one could also say that
for a given value of the density of the accelerating component
(a cosmological constant in the standard case or the bulk-induced
term in the DGP model) the DGP model requires a higher matter
density in order to fit the supernova data. Note that the two and
three sigma likelihood contours are quite close to each other, and
relatively distant from the one sigma contour. This indicates that
with the currently available data there is an elongated `degenerate
best-fit plateau', and beyond this plateau the likelihood drops
quite abruptly.

In any case, just by looking at Fig. \ref{fig1} one might think that
there is a rather comfortable range of matter densities which would
give models in agreement with observation. However this is not the case 
as there are other cosmological constraints that must be met.
In particular, the most recent CMB
data \citep{boom1,boom2,boom3,maxi1,maxi2,maxi3,dasi1,dasi2} gives
a strong indication that the universe is spatially flat or very
nearly so. The current constraint is
\begin{equation}
\label{cmbfix}
\Omega_{tot}=1.00\pm0.05\,;
\end{equation}
note that only fairly weak priors are needed to derive this constraint
(refer to the original CMB papers for the analysis details).
Combining this with the supernova analysis this leaves a much smaller
range of allowed models. At the $68\%$ confidence
level, the allowed range of matter densities is approximately
\begin{equation}
\label{dens1sigma}
\Omega_m=0.20\pm0.05\,,
\end{equation}
while at the $99\%$ confidence
level it is
\begin{equation}
\label{dens3sigma}
\Omega_m=0.2\pm0.1\,.
\end{equation}
Note that this result is quite robust---for example, we have checked that
it is unchanged if the likelihood analysis is restricted {\em ab
initio} to flat universes.

The final piece of observational evidence that we shall use are
dynamical measurements of the total mass density---see \citet{turner1,turner2}
for a discussion of the state-of-the-art. In particular the ratio of
baryons to the total mass in clusters has been determined using both
X-ray measurements \citep{mohr} and Sunyaev-Zel'dovich
measurements \citep{carlstrom}. One respectively obtains
\begin{equation}
\label{fraction1}
f_x=(0.075\pm0.007)h^{-3/2}\,
\end{equation}
from X-ray measurements, and
\begin{equation}
\label{fraction2}
f_{sz}=(0.079\pm0.010)h^{-1}\,
\end{equation}
from the Sunyaev-Zel'dovich effect. In both cases, various dozens of
sources have been used in the analysis.
If one assumes that clusters are a fair sample of the matter content
of the universe (which is very reasonable given their large
size) and uses the latest value of the baryon density at
nucleosynthesis \citep{bbn}
\begin{equation}
\label{baryon0}
\Omega_b=(0.020\pm0.002)h^{-2}\,,
\end{equation}
together with the value of the (re-scaled) Hubble constant $h$
obtained by the HST Key Project \citep{hst},
\begin{equation}
\label{littleh1}
h=0.72\pm0.08\,,
\end{equation}
one finally obtains the (rather conservative) estimate
\begin{equation}
\label{baryonfinal}
\Omega_m=0.35\pm0.07\,.
\end{equation}
More recently, there have been claims of an even narrower (though
perhaps slightly optimistic) range \citep{newomega} at the one sigma
level
\begin{equation}
\label{baryonfinalnew}
\Omega_m=0.330\pm0.035\,.
\end{equation}
Note also that there are various other sources of supporting evidence
that are consistent with the above value, including studies
of the evolution of cluster abundances with redshift, measurements of the
power spectrum of large-scale structure (such as the recent preliminary
2dF results \citep{tdf}), analyses of measured peculiar
velocities as they relate to the observed matter distribution,
and observations of the
outflow of material from voids. A discussion of the assumptions and
techniques of each method can be found in \citet{turner1,turner2}.

Hence the DGP model's proposal for the acceleration of the universe
requires a value of the matter density that is inconsistent, at least at
the two sigma level, with the observationally estimated
matter density of the universe. Together with the fact that
gravity becomes weaker on large enough scales,
this presents a serious problem. Note that
if the required mass was larger than the standard case, one could perhaps
argue that there was some matter in an yet undetected form. Indeed,
the fact that in the DGP model gravity becomes weaker on large enough
scales could then be used to obtain a relatively simple explanation.
However, since the observationally acceptable range of masses is lower
than the standard model, no explanation of this kind is possible.
In any case, one should recall that the evolution of the universe
should be as standard in the DGP model up to very recent times, eg in
what concerns the Friedmann equation for example---which also places
strong constraints on any attempts to `get rid of' some of the matter.

\begin{figure}
\plotone{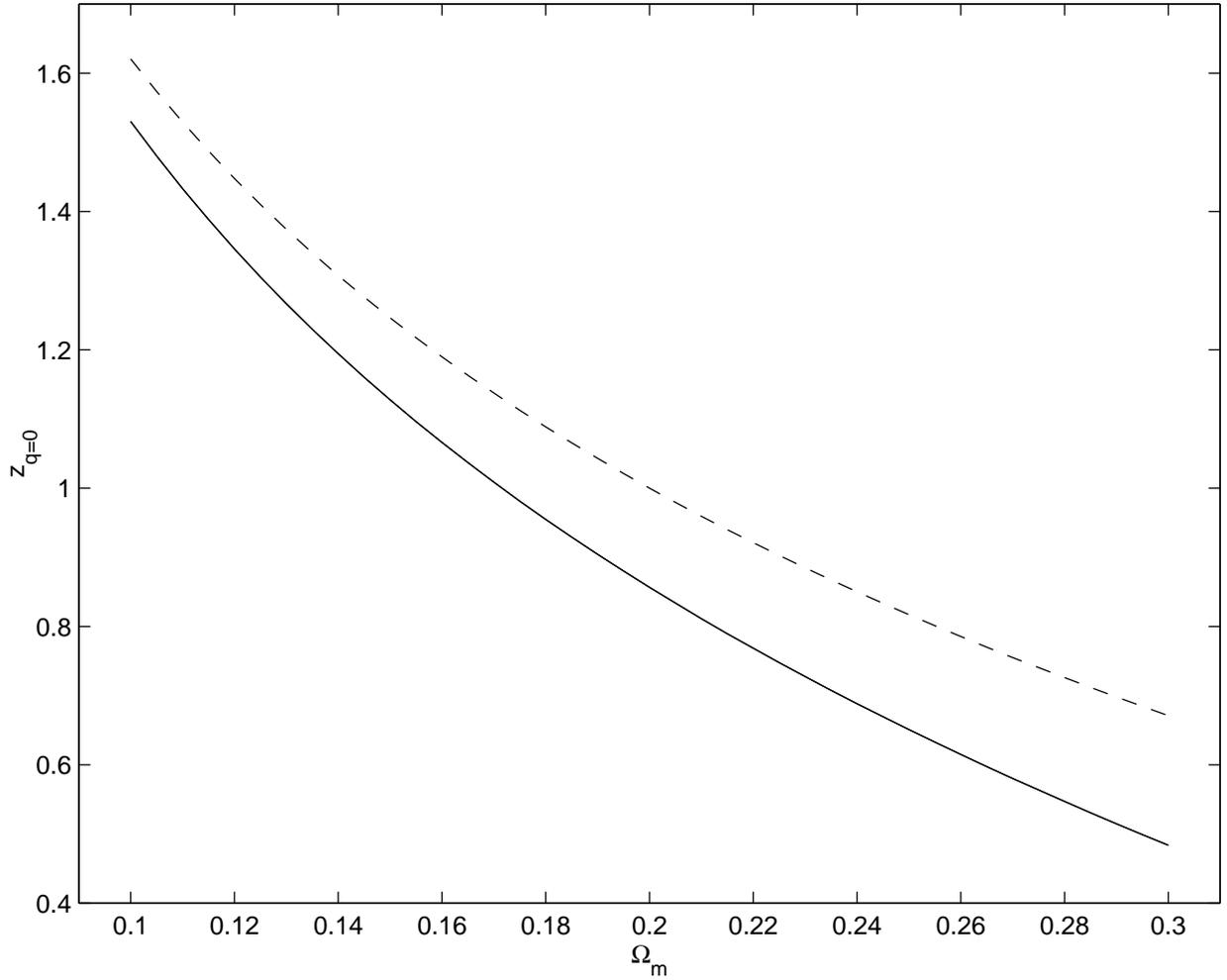}
\caption{The redshift of `turnaround' for which the deceleration
parameter vanishes, as a function of the matter density in the DGP
models, for the values of the matter density that are consistent
(up to $99 \%$ confidence level) with type Ia supernova and CMB data
(solid line), and the analogous quantity in the standard cosmological
model (dashed line). In both cases $\Omega_{tot}=1$. \label{fig3}}
\end{figure}

And finally, there is yet another hurdle for this model to overcome.
In (\ref{qvanishes}) we derived the redshift at which the universe
switches from deceleration to acceleration, for the case of a flat
universe. We plot this redshift, for the range of matter densities
given by (\ref{dens3sigma}), in Fig. \ref{fig3}, together with
the analogous curve for the standard model. The problem is now
apparent: for the specified range of matter densities, the redshift
of turnaround {\em decreases as the matter density increases}. On the
other hand, for a given value of the turnaround redshift, the required
matter density is always lower with a bulk-induced
term replacing a cosmological constant (the solid curve in
Fig. \ref{fig3}) than in the standard model with a cosmological
constant (the dashed curve). Now,
the latest supernova data \citep{perl1,perl2,riess,newq1,newq2,our}
indicates that the universe switched from deceleration to acceleration
at a redshift in the interval (at one-sigma level)
\begin{equation}
\label{switchz}
0.6<z_{q=0}<1.7\,.
\end{equation}
Note that for values of the matter density close to the upper limit
$\Omega_m\sim0.3$ the predicted redshift of turnaround is already smaller
than this range.

This, therefore, is the dilemma of these models. On
the assumption of a flat universe, a very low matter density is needed
so that acceleration starts early enough. This is in fact confirmed
by the simulation of the supernova analysis for a SNAP-class
dataset, which is shown in Fig. \ref{fig2}: closed models are still
favoured, though flat ones are still possible at around the two sigma
level. However the range of possible matter densities is significantly
reduced. Note that in generating the SNAP dataset we have assumed a
standard FRW universe with $\Omega_m=0.3$ and $\Omega_\Lambda=0.7$.
Hence, according
to our discussion above, if we fit that dataset to the 
accelerating DGP model then
the preferred value of the matter density will come out lower. This is
a trivial consequence of the fact that the type Ia supernova analysis
method is basically a  cosmological `accelerometer'.
The point is that, even with the data available today,
such low values are already strongly disfavoured by dynamical measurements of
the total mass density in the universe.

On the other hand, even if one would be willing to admit that such
values were allowed on the grounds of dynamical measurements alone
(implying a much smaller value of dark matter than in the standard
model), they are expected to run into serious difficulties when it comes to
density fluctuation growth and the evolution of large-scale structures
(which can now be probed much beyond Mpc scales both by direct surveys
and through gravitational lensing), again because of the
weaker gravity on large enough scales. For the DGP scenario to be
viable the characteristic scale would be of order
$r_c\sim H_0^{-1}$, but obviously the effects of weaker gravity
would be felt on smaller scales than this.
Indeed this point has already
been made on rather general grounds (though only for the case
of sub-horizon modes) by \citet{uzan}, and we shall return to it
in more detail elsewhere.

\section{Conclusions}

In this paper we have considered the cosmological consequences of the
brane world model of \citet{dgp}, and its proposed alternative
explanation for the current acceleration of the universe. We have
shown that, contrary to recent claims \citep{ddg}, this proposal
is already strongly disfavoured by existing cosmological datasets,
at least at the two sigma level.
In order to be consistent with CMB and supernova data one would
need a very low matter density $\Omega_m\sim0.2$. Even if this
was allowed by dynamical measurements (such as cluster data),
such a low density ( and hence such a small amount of dark
matter) together with the fact that gravity is weaker on large
enough scales would make it difficult to produce
a consistent structure formation scenario.

The lesson to be learned from this exercise is twofold. Firstly, no matter
how interesting or mathematically clever one's favourite particle physics
model of the universe might be, the first hurdle towards credibility
consists in deriving falsifiable cosmological predictions from it.
And secondly, the currently available cosmological observations are
already powerful enough to impose tight constraints on a wide
range of possible models, especially when various cosmological datasets
are combined---which is a sign that the era of precision cosmology
has indeed started. We hope that other brane world scenarios can be brought
into the realm of cosmological testability in the near future.

\acknowledgments

C. M. is grateful to the organizers and participants of the Extra
Dimensions Workshop at Coll\`ege de France, Paris, for many enlightening
seminars and discussions.
C.\ M.\ is funded by FCT under  ``Programa PRAXIS XXI'' (grant no.\ 
FMRH/BPD/1600/2000). We thank `Centro de Astrof{\' \i}sica da Universidade
do Porto' (CAUP) for the facilities provided.

\clearpage

\clearpage

\clearpage


\begin{thebibliography}{}
\bibitem[Arkani-Hamed, Dimopoulos \& Dvali(1998)]{orig1}
Arkani-Hamed, N., Dimopoulos, S., and Dvali, G. 1998, Phys. Lett. B429, 263
\bibitem[Avelino, de Carvalho \& Martins(2001)]{our}
Avelino, P.P., de Carvalho, J.P.M., and Martins, C.J.A.P. 2001,
Phys. Rev. D, in press (see astro-ph/0103075)
\bibitem[Balbi et al.(2000)]{maxi2}
Balbi, A. {\em et al.} 2001, \apjl 545, 5
\bibitem[Behnke et al.(2001)]{behnke}
Behnke, D., Blaschke, D., Pervushin, V.N., and Proskurin, D. 2001,
gr-qc/0102039
\bibitem[Bin\'etruy et al.(2000)]{orig3}
Bin\'etruy, P., Deffayet, C., Ellwanger, U., and Langlois, D. 2000,
Phys. Lett. B477, 285
\bibitem[Burles, Nollett \& Turner(2000)]{bbn}
Burles, S., Nollett, K.M., and Turner, M.S. 2000, astro-ph/0010171
\bibitem[Carlstrom et al.(1999)]{carlstrom}
Carlstrom, J. {\it et al.} 1999, astro-ph/9905255
\bibitem[de Bernardis et al.(2000)]{boom1}
de Bernardis, P. {\em et al.} 2000, Nature 404, 955
\bibitem[Deffayet(2001)]{def}
Deffayet, C. 2001, Phys. Lett. B502, 199
\bibitem[Deffayet, Dvali \& Gabadadze(2001)]{ddg}
Deffayet, C., Dvali, G., and Gabadadze, G. 2001, astro-ph/0105068
\bibitem[Dev, Sethi \& Lohiya(2000)]{dev}
Dev, A., Sethi, M., and Lohiya, D. 2000, astro-ph/0008193
\bibitem[Dick(2001)]{rdick}
Dick, R. 2001, hep-th/0105320
\bibitem[Dvali, Gabadadze \& Porrati(2000)]{dgp}
Dvali, G., Gabadadze, G., and M. Porrati, M. 2000, Phys. Lett. B485, 208
\bibitem[Freedman et al.(2000)]{hst}
Freedman, W.L. {\it et al.} 2000, astro-ph/0012376
\bibitem[Halverson et al.(2001)]{dasi1}
Halverson, N.W. {\em et al.} 2001, astro-ph/0104489
\bibitem[Hanany et al.(2000)]{maxi1}
Hanany, S. {\em et al.} 2001, \apjl 545, 1
\bibitem[Lange et al.(2001)]{boom2}
Lange, A.E. {\em et al.} 2001, Phys. Rev. D63, 042001
\bibitem[Maartens(2001)]{review1}
Maartens, R. 2001,  gr-qc/0101059
\bibitem[Mannheim(2001)]{mann}
Mannheim, P. 2001, astro-ph/0104022
\bibitem[Mohr et al.(1999)]{mohr}
Mohr, J. {\it et al.} 1999, \apj 517, 627
\bibitem[Netterfield et al.(2001)]{boom3}
Netterfield, C.B. {\em et al.} 2001, astro-ph/0104460
\bibitem[Percival et al.(2001)]{tdf}
Percival, W.J. {\em et al.} 20001, astro-ph/0105232
\bibitem[Perlmutter et al.(1997)]{perl1}
Perlmutter, S. {\em et al.} 1997, \apj 483, 565
\bibitem[Perlmutter et al.(1999)]{perl2}
Perlmutter, S. {\em et al.} 1999, \apj 517, 565
\bibitem[Pryke et al.(2001)]{dasi2}
Pryke, C. {\em et al.} 2001, astro-ph/0104490
\bibitem[Randall \& Sundrum(1999)]{orig2}
Randall, L., and Sundrum, R. 1999, Phys. Rev. Lett. 83, 3370
\bibitem[Riess et al.(1998)]{riess}
Riess, A. {\em et al.} 1998, \aj 116, 1009
\bibitem[Riess et al.(2001)]{newq1}
Riess, A.G. {\it et al.} 2001, astro-ph/0104455
\bibitem[Rubakov(2001)]{review2}
Rubakov, V. 2001, hep-ph/0104152
\bibitem[Stompor et al.(2001)]{maxi3}
Stompor, R. {\em et al.} 2001, astro-ph/0105062
\bibitem[Turner(2000a)]{turner1}
Turner, M.S. 2000, Physica Scripta T85, 210
\bibitem[Turner(2000b)]{turner2}
Turner, M.S. 2000, Physics Reports 333, 619
\bibitem[Turner(2001)]{newomega}
Turner, M.S. 2001, astro-ph/0106035
\bibitem[Turner \& Riess(2001)]{newq2}
Turner, M.S., and Riess, A.G. 2001, astro-ph/0106051
\bibitem[Uzan \& Bernardeau(2000)]{uzan}
Uzan, J.-P., and Bernardeau, F. 2000, hep-th/0012011
\bibitem[Wang \& Garnavich(2001)]{garn}
Wang, Y., and Garnavich, P. 2001, \apj 552, 445
\bibitem[Wang et al.(2000)]{wang}
Wang, Y. {\em et al.} 2000, \apj 536, 531
\bibitem[Weller \& Albrecht(2001)]{weller}
Weller, J., and Albrecht, A. 2001, Phys. Rev. Lett. 86, 1939
\end{thebibliography}
\end{document}